\documentclass[twocolumn]{article}
\advance\topmargin-1in\advance\oddsidemargin-1in
\evensidemargin\oddsidemargin

\makeatletter
\def\@normalsize{\@setsize\normalsize{12pt}\xpt\@xpt
\abovedisplayskip 10pt plus2pt minus5pt\belowdisplayskip
\abovedisplayskip \abovedisplayshortskip \z@
plus3pt\belowdisplayshortskip 6pt plus3pt
minus3pt\let\@listi\@listI}

\def\section{\@startsection {section}{1}{\z@}{20pt plus 2pt minus 2pt}
{8pt plus 2pt minus 2pt}{\centering\normalsize\sc
\edef\@svsec{\thesection.\ }}}
\def\thesection{\Roman{section}}

\def\subsection{\@startsection {subsection}{2}{\z@}{16pt plus 2pt minus 2pt}
{6pt plus 2pt minus 2pt}{\normalsize\sl
\edef\@svsec{\thesubsection.\ }}}
\def\thesubsection{\Alph{subsection}}

\long\def\@makecaption#1#2{ \vskip10pt\begin{center} #1 #2
\end{center}\par\vskip 1pt}
\def\fnum@figure{\raggedright{\footnotesize Fig. \thefigure }.%
\footnotesize}
\def\fnum@table{\footnotesize TABLE \thetable\\\footnotesize\sc}
\def\thetable{\Roman{table}}

\usepackage{algorithm}
\usepackage{algorithmic}

\usepackage{graphicx}
\graphicspath{{./}{./figs/}}
\makeatother

\begin{document}
\date{}

\title{\Large\textbf{Floorplanning and Topology Generation for Application-Specific Network-on-Chip}}

\author{
 Bei Yu, \ \ Sheqin Dong\\
 Department of Computer Science \& Technology \\
 TNList\footnote{Tsinghua National Laboratory for Information Science and Technology}\\
 Tsinghua University, Beijing, China\\
 \and
 Song Chen, \ \ Satoshi GOTO\\
 Graduate School of IPS\\
 Waseda University, Kitakyushu, Japan\\
 }

 \maketitle

{\small\textbf{Abstract--- Network-on-Chip(NoC) architectures have
been proposed as a promising alternative to classical bus-based
communication architectures. In this paper, we propose a two phases
framework to solve application-specific NoCs topology generation
problem. At floorplanning phase, we carry out partition driven
floorplanning. At post-floorplanning phase, a heuristic method and a
min-cost max-flow algorithm is used to insert switches and network
interfaces. Finally, we allocate paths to minimize power
consumption. The experimental results show our algorithm is
effective for power saving.}}

\section{Introduction}
Network-on-Chip(NoC) architectures have been proposed as a promising
alternative to classical bus-based and point-to-point communication
architectures when the CMOS technology entered the nanometer era
\cite{ieee02NOC,ieee05Bertozzi,ieee09Marculescu}. In NoCs, the
communication among various cores is achieved by on-chip
micro-networks components(such as switch and network interface)
instead of the traditional non-scalable buses.

Comparing with bus-based architectures, NoCs have better modularity
and design predictability. Besides, the NoC approach offers lower
power consumption and greater scalability.

NoCs can be designed as regular or application-specific network
topologies. For regular Noc topology design, some existing NoC
solutions assume a mesh-based NoC architecture
\cite{ASPDAC03HU,DATE04MURALI}, and their focus is on the mapping
problem. For application-specific topology design, the design
challenges are different in terms of irregular core sizes, various
core locations, and different communication flow requirements
\cite{ieee06Srinivasan,ICCAD06MURALI,ASPDAC09MURALI,ASPDAC08Chan,ASPDAC08Yan}.
Most SoCs are typically composed of heterogeneous cores and the core
sizes are highly non-uniform. An application-specific NoCs
architecture with structured wiring, which satisfies the design
objectives and constraints is more appropriate. In this paper, we
focus on synthesis problem of application-specific NoCs
architecture.

Network components, such as switches and network interfaces(NI),
consume area and power. The area consumption of these network
components should be considered during topology generation.
Besides, power efficiency is one of the most important concerns in
NoCs architecture design. Many characteristics influence NoCs power
consumption: total wirelength; communication flow distributions and
path choosing. In this paper, we propose a methodology to design the
best topology that is minimize power consumption of interconnects
and network components.

%*********************************************************
%                   Previous Works
There are a number of works addressing NoCs topology generation. In
\cite{ieee06Srinivasan}, a novel NoC topology generation algorithms
were presented, however their solutions only consider topologies
based on a slicing structure where switch locations are restricted
to corners of cores. In \cite{ICCAD06MURALI}, Murali et al. proposed
a two steps topology generation procedure using a min-cut
partitioner to cluster highly communicating cores on the same switch
and a path allocation algorithm to connect the clusters together. In
\cite{ASPDAC08Chan}, Chan et.al. presented an iterative refinement
strategy to generate an optimized NoC topology that supports both
packet-switched networks and point to point connections.

In most of the previous works, system-level floorplanning tool is
used only estimates the area and the wire lengths. Partition is
carried out at pre-floorplanning, so physical information such as
the distances among cores are not able to be taken into account.
Besides, area of switches and network interfaces are not consider
during topology generation.

In this paper, we integrate partition into floorplanning to make use
of physical information such as the length of interconnects among
cores. At post-floorplanning optimization, a heuristic method is
used to insert switches and a min-cost max-flow algorithm is used to
insert network interfaces. Finally, we allocate paths to minimize
power consumption.

%the area consumption of switches and NI are taken into account.

The remainder of this paper is organized as follows. Section 2
defines the partition driven floorplanning problem. Section 3
presents our algorithm flow. Section 4 reports our experimental
results. At last, Section 5 concludes this paper.

\section{Problem Formulation}
\newtheorem{define}{Definition}
\newtheorem{Theorem}{Theorem}
\newtheorem{problem}{Problem}

\begin{figure}[tb]
\centering
\includegraphics[width=0.48\textwidth]{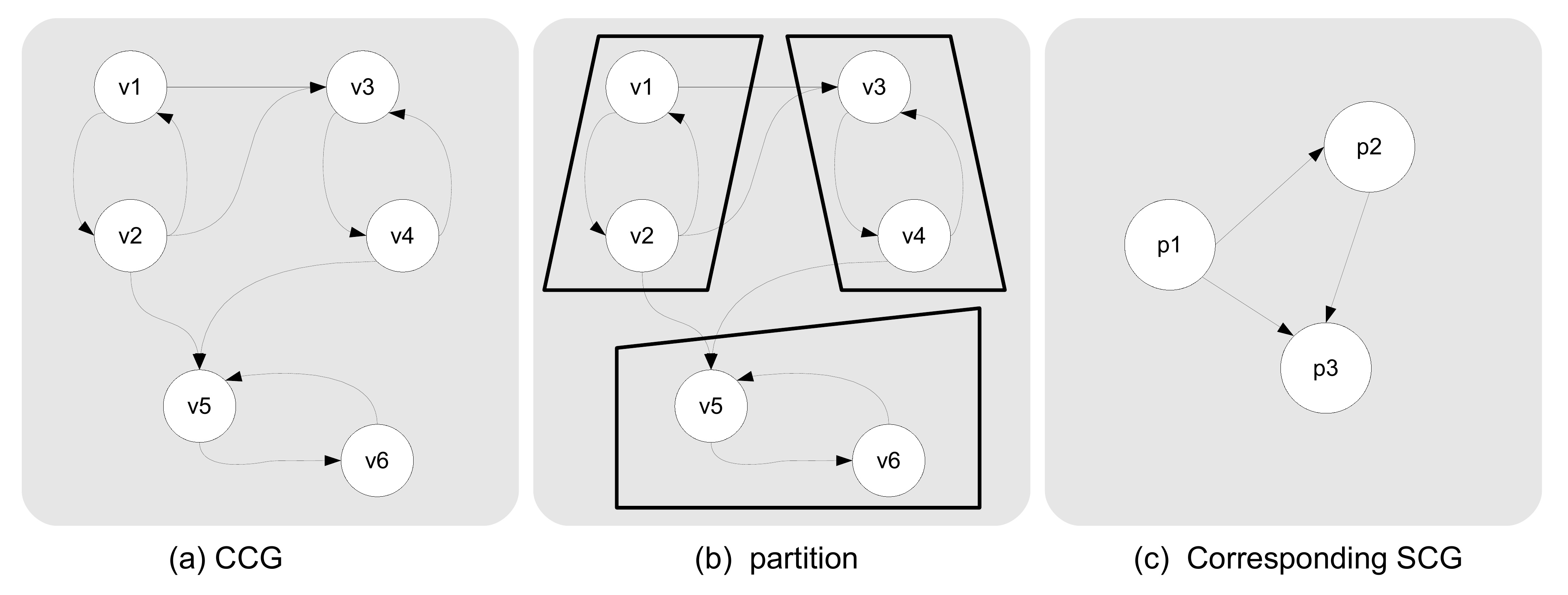}.
\caption{CCG and SCG examples.~(a)A simple CCG.~(b)CCG is
partitioned based on communication requirements and related
positions.~(c)Corresponding SCG.} \label{ccg}
\end{figure}

\begin{define}[Core Communication Graph(CCG)]
The core communication graph is a directed graph, $\bar G=(\bar V,
\bar E)$ with each vertex $v_i \in \bar V$ representing a core and
the edge $e_{ij}$ representing the communication requirement between
the core $v_i$ and $v_j$. The weight of edge $e_{ij}$ is denoted as
$w_{ij}$.
\end{define}

\begin{define}[Switch Communication Graph(SCG)]
The switch communication graph is a directed graph, $G=(V, E)$ with
each vertex $v_i \in V$ representing a switch, and the directed edge
$e_{ij} = \{v_i, v_j\} \in E$ denotes a communication trace from
$v_i$ to $v_j$.
\end{define}

A simple CCG with six cores is shown in Fig.\ref{ccg}(a). After
partition, corresponding SCG with three switches are generated as
shown in Fig.\ref{ccg}(c).

\begin{define}[Cluster Bounding Resource]
The cluster bounding resource of a cluster is evaluated by the half
perimeter wirelength of the minimal bounding box enclosing the
cluster.
\end{define}

\begin{problem}{\rm \textbf{(NoCs Topology Generation)}}
The topology generation problem can be defined as follows:
\textbf{given} a set of $n$ cores $C=\{c_1, c_2, \dots, c_n\}$, a
switches number constraint $m$, a core communication graph(CCG) and
network components power model, \textbf{find} an NoC topology that
satisfies several objectives: minimize area consumption of cores and
network components($m$ switches and $n$ network interfaces);
minimize the communication energy.

\end{problem}

Cores with more communication requirements are incline to be
assigned into same cluster to minimize communication energy.
Relative positions of cores should be considered during partition to
minimize area consumption. Besides, positions of network components,
such as switches and network components, should be taken into
account to minimize interconnect length. Finally, the actual
physical connections between switches are established to find paths
minimizing traffic flows energy across the switches.

\section{Topology Synthesis Algorithm}

\begin{figure}[tb]
\centering
\includegraphics[width=0.44\textwidth]{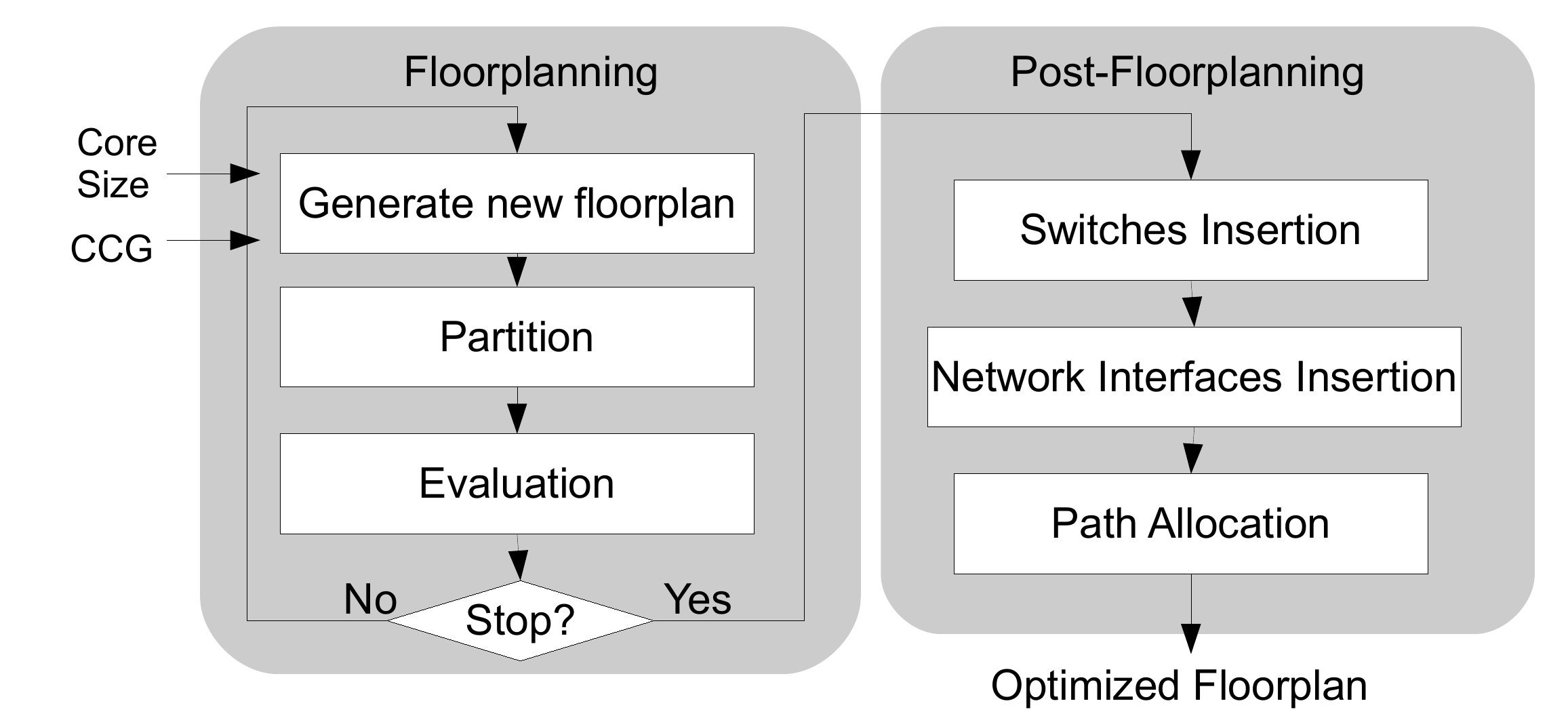}.
\caption{Topology Synthesis Algorithm Overall} \label{overview}
\end{figure}

As shown in Fig.\ref{overview}, the algorithm flow consists of two
phases: (I)partition driven floorplanning, (II)post-floorplanning
optimization.

In Phase I, we integrate partition into floorplanning. When generate
a new packing, we carry out partition to assign each core into one
cluster. Partition should consider not only communication
requirements among cores but also physical information of cores.

In Phase II, in switches insertion, a heuristic method is adopted to
calculate every switch's position in white space. In network
interfaces insertion, we present a Min-Cost Max-Flow based method to
insert each NI in white space. Finally, an effective incremental
path allocation method is proposed to minimize power consumption.

\subsection{Partition Driven Floorplanning}
Traditionally, floorplanning tool is only used to evaluate the wire
lengths between each cores and switches. And partition is carried
out before floorplanning, so physical information such as the
distances among modules are not able to be taken into account during
partition.

In this paper, we integrate partition into floorplanning phase.
During floorplanning, after generating a new chip floorplan, we can
estimate the interconnect length between module i and module j,
denoted as $len_{ij}$. Given core communication graph(CCG) and
switches number constraint $m$, partition assign cores into $m$
min-cut clusters. Those cores with larger communication requirements
and less distances are assigned to the same cluster and hence use
the same switch for communication. On the one hand, cores with
larger communication requirements are more incline to cluster
together to minimize interconnect power consumption. On the other
hand, cores with less distances should be cluster to minimize
cluster bounding resource.

The partitioning is done in such a way that the edges of the graph
that are cut between the partitions have lower weights than the
edges that are within a partition and the number of vertices
assigned to each partition is almost the same. In partition, we
define new edge weight $w_{ij}'$ in CCG:
\begin{equation}
  w_{ij}' = \alpha_w \times \frac{w_{ij}}{max\_w} + \alpha_d \times
  \frac{mean\_dis}{dis_{ij}}
\end{equation}
where $w_{ij}$ denotes communication requirement between core $i$
and core $j$, $dis_{ij}$ denotes distance between core $i$ and $j$,
$max\_w$ is the maximum communication requirement over all flows and
$mean\_dis$ is average distance among cores.

During floorplanning, we use CBL\cite{CBL04} to represent every
floorplan generated. CBL is a topological representation dissecting
the chip into rectangular rooms. The cost function in simulated
annealing is:
\begin{equation}
\Phi = \lambda_AA + \lambda_FF + \lambda_RR
\end{equation}
where $A$ represent the floorplan area; $F$ represents the total
communication amount between clusters; and $R$ represents the sum of
all cluster bounding resources. The parameters $\lambda_A$,
$\lambda_F$ and $\lambda_R$ can be used to adjust the relative
weighting between the contributing factors.

\begin{figure}[tb]
\centering
\includegraphics[width=0.44\textwidth]{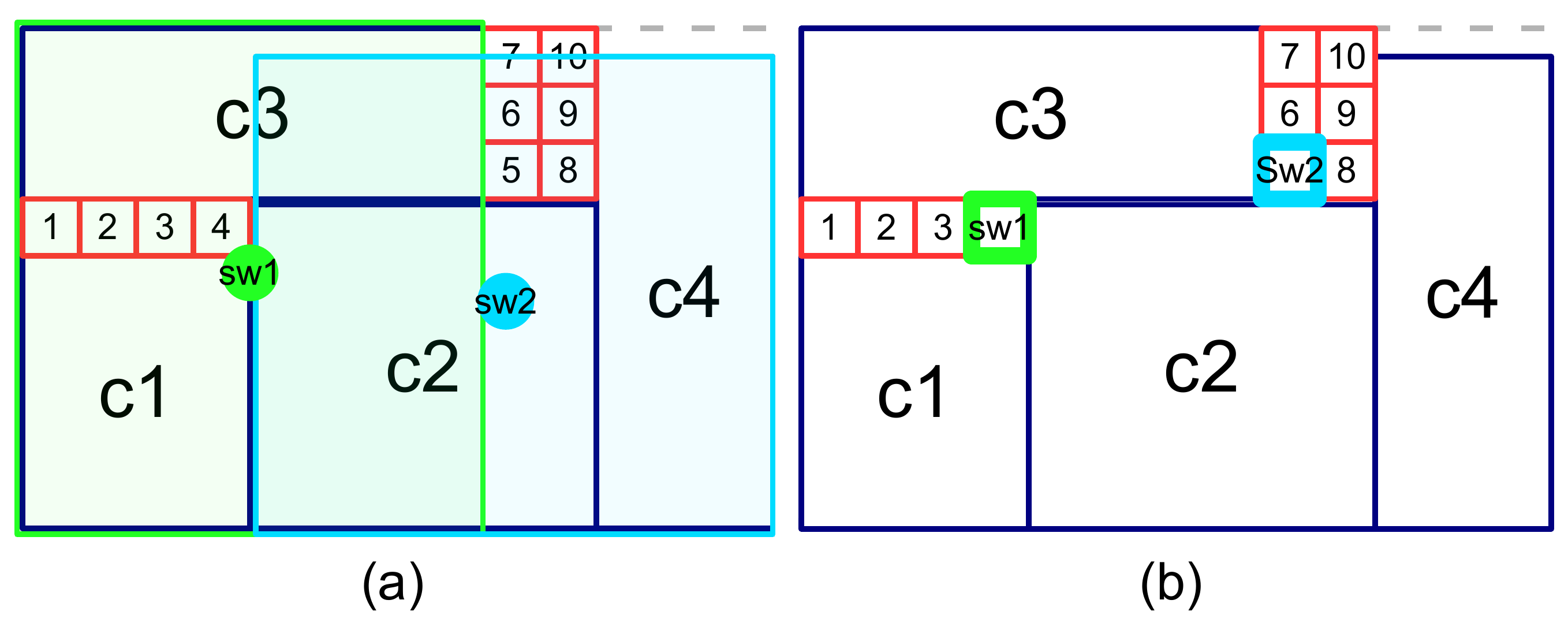}.
\caption{~A floorplan with four cores, in which white spaces are
divided into grids(label from $1$ to $10$).~(a)Core c1 and c3 are
partitioned into one cluster and c2 and c4 are partitioned into
another cluster. Two dots are initial positions of two
switches.~(b)Switches are assigned to grids one by one and finally
two switches $sw1$ and $sw2$ have decided their positions.}
\label{grids}
\end{figure}

\subsection{Switches Insertion}

Once a floorplan with $m$ clusters $P=\{p_1, p_2, \dots p_m\}$ is
obtained, the next step is to find the latency and power consumption
on the wires. In order to do this, the position of the switches
needs to be determined. Each cluster has one switch and
communication among clusters are through switches. We denote the set
of switches as $SW = \{sw_1, sw2, \dots sw_{m}\}$, and switch $sw_k$
belongs to cluster $p_k$. Due to the restriction that switches
cannot be placed on a core, the location must be within a white
space.

We partition the dead space into grids and each grid provides sites
for switches insertion. Then a heuristic method is proposed to
insert each switch into one grid(as shown in Fig. \ref{grids}).

The minimal bounding box enclosing cluster $p_k$ is defined as
$B_k$. For switch $sw_k$, its candidate grids are the free grids
inside $B_k$. For example, in Fig. \ref{grids}(a), cluster $p_1$
includes core c1 and core c2, and switch $sw1$'s candidate grids are
label from $1$ to $4$. Switch $sw2$'s candidate grids are those
label $5,6,8,9$. Initially, each switch $sw_k$ is located in the
center of cluster's bounding box.

For switch $sw_k$, its communication requirement is define as
follow:
\begin{equation}
  flow_k = \sum_{i, j} w_{ij}, \forall e_{ij}\in
\bar E~\&~i\in p_k~\&~j \notin p_k
\end{equation}
where $i\in p_k$ means core $i$ is assigned to cluster $p_k$.

We sort switches by their communication requirements, and assign
each switch into one of its candidate grids one by one. If one free
grid(label $g$) is candidate grid of switch $sw_k$, then the
insertion cost $Cost_{gk}$ is defined as follow:
\begin{equation}
  Cost_{gk} = \sum_{i,j}w_{ij}\times
  (dis_{gi}+dis_{gj}), \forall e_{ij} \in \bar E~\&~i\in p_k~\&~j \notin p_k
\end{equation}
where $dis_{gi}$ is the distance from grid $g$ to core $i$. Each
switch chooses one of the candidate grids with lest insertion cost
to insert. As shown in Fig. \ref{grids}(b), $sw1$ inserts into grid
4 and $sw2$ inserts into grid 5.

\subsection{Network Interfaces Insertion}

\begin{figure}[tb]
\centering
\includegraphics[width=0.44\textwidth]{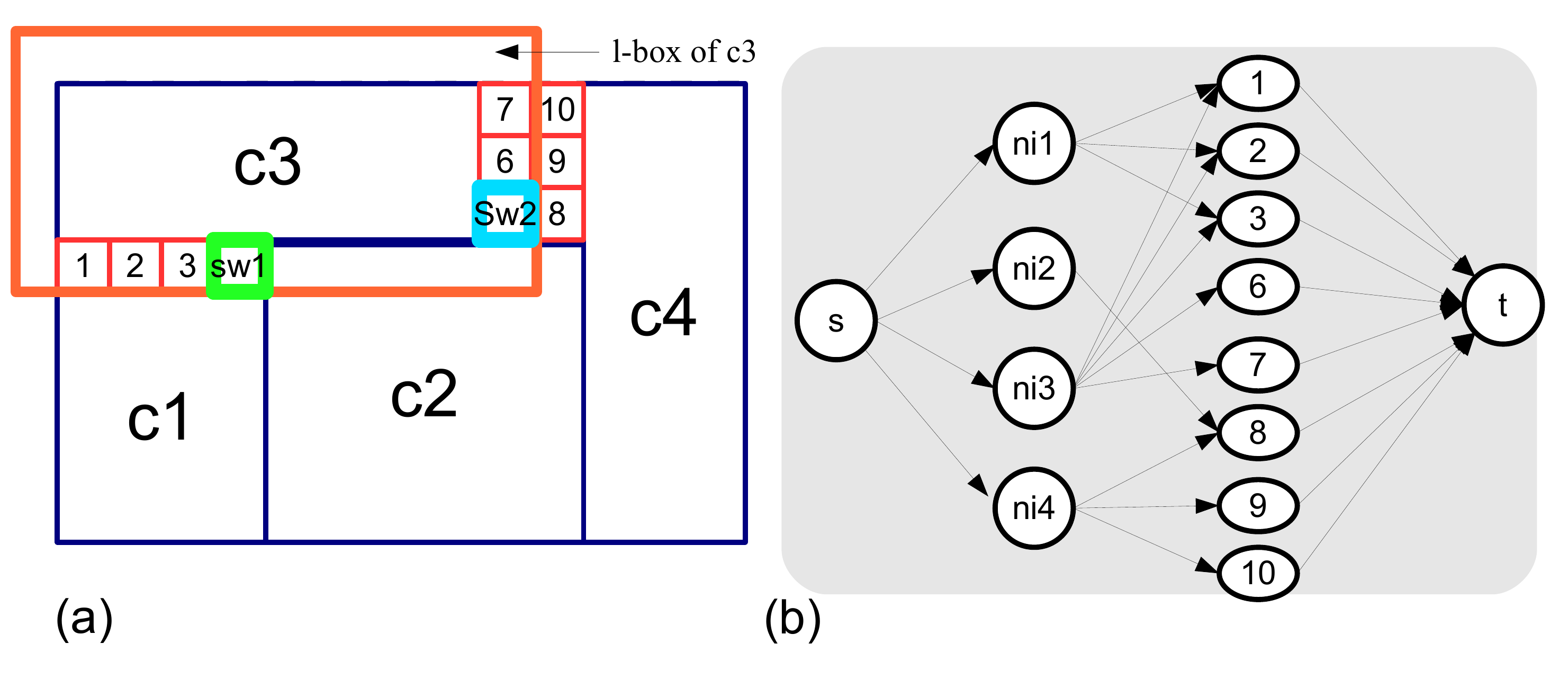}
\caption{~A simple example of network interfaces insertion.~(a)When
$l$ sets as the width of grid, l-box of core $c3$ includes five free
grids(label $1, 2, 3, 6, 7$). ~(b) Corresponding network flow
model.} \label{network}
\end{figure}

After switches insertion, every switch is assigned a grid in white
space. Then we carry out minimum cost flow based network interfaces
insertion to assign each NI into one grid. We define set of Network
Interfaces as $NI = \{ni_1, ni_2, \dots ni_n\}$, where $n$ is number
of cores. Each core $c_k$ needs one network interface $ni_k$ to
connect to switch.

\begin{define}[$l$-bounding box]Given a core $c_k$, whose
width is $wid_k$ and height is $hei_k$. The $l$-bounding box of
$c_k$ is $Bl_k$, which has the same centric position. Besides, width
of $Bl_k$ is $(wid_k+2 \times l)$ and height is $(hei_k+2 \times l)$
(as shown in Fig.\ref{network}(a)).
\end{define}

For each core $c_k$, we construct its $l$-bounding box. The free
grids in the $l$-bounding box are $c_k$'s candidate grids, denoted
as $CG_k$.

We construct a network graph $G^* = (V^*, E^*)$, and then use a
min-cost max-flow algorithm to determine which grid each network
interface belong to. A simple example is shown in Fig.\ref{network}.

\begin{flushleft}
\begin{itemize}
\item
  $V^* = \{s, t\}\cup NI \cup Grids$.
\item
  $E^* = \{ (s, ni_k)|ni_k\in NI\} \cup \{ ( ni_k, g_j) | \forall g_j\in CG_k \} \cup \{( g_j, t)| g_j \in Grids \}$.
\item
  Capacities: $C(s,ni_k)=1, C(ni_k, g_j)=1, C(r_j,
  t ) = 1$.
\item
  Cost: $F(s, ni_k)=0, F(g_j, t)=0; F(ni_k, g_j)=F_{kj}$.
\end{itemize}
\end{flushleft}
where $F_{kj}$ equals to distance from grid $j$ to switch $sw_k$.

Network Interfaces insertion can be solved effectively by minimum
cost flow algorithm(run in polynomial time\cite{book:flow}).

\subsection{Energy Aware Path Allocation}
After switches insertion, we use dynamic programming based method
for path allocation to minimize power assumption.

Given switch communication graph(SCG) $G=(V, E)$ representing
communication requirement among switches. The communication
requirement of $e_{ij}\in E$ denoted as $ws_{ij}$:
\begin{equation}
  ws_{ij}=\sum_{\forall a\in p_i}\sum_{\forall b\in
  p_j}(w_{ab}+w_{ba})
\end{equation}
where $p_i$ is cluster $i$ and $w_{ab}$ is communication requirement
from core $c_a$ to core $c_b$.

We denote nodes in SCG as $v_1, v_2, \dots, v_m$, where $m$ is the
number of switches. We assume SCG only exists directed edge $e_{ij}$
that $i < j$ because $e_{ij}$ represents both communication from
switch $sw_i$ to $sw_j$ and $sw_j$ to $sw_i$.

\begin{table}[bt]\label{table:path}
\centering \caption{Notation used in Path Allocation}
\begin{tabular}{|l|l|}
 \hline \hline
 $t_{ij}$ & power consumption to connect $e_{ij}$.\\
 \hline
 $Pre(i)$ & $\{v_k|\forall v_k \in V ~\&~ e_{ki}\in E\}$ \\
 \hline
 $Post(i)$ & $\{v_k|\forall v_k \in V ~\&~e_{ik}\in E\}$ \\
 \hline
 $dis_e(i, j, d)$ & minimum distance from node $v_i$ to $v_d$\\
               & while edge $e_{ij}$ is used. \\
 \hline
 $dis_n(i, d)$    & minimum distance from node $v_i$ to $v_d$.\\
 \hline
 $path(i, d)$  & denote which node $v_i$ connect to go to $v_d$.\\
 \hline \hline
\end{tabular}
\end{table}

As shown in Table I, we define set $Pre(i)$ as $v_i$'s front-end
nodes and $Post(i)$ as $v_i$'s back-end nodes. We also define two
kind of distance $dis_e(i, j, d)$ and $dis_n(i, d)$. Besides,
$path(i, d)$ denotes which node $v_i$ should connect to go to $v_d$.
We use the following ways to solve $dis_e$, $dis_n$ and $path$:

 \begin{equation}
   dis_e(i, j, d) = \left \{
     \begin{array}{ll}
       t_{id},     & j=d ~\&~ i\in Pre(d) \\
       t_{ij}+dis_n(j, d), & otherwise \\
      \end{array}
   \right.
 \end{equation}
 \begin{equation}
   dis_n(i, d) = \left \{
     \begin{array}{ll}
       0,     & i=d\\
       min_k~ dis_e(i, k, d), & \forall k \in Post(i) \\
      \end{array}
   \right.
 \end{equation}
 \begin{equation}
   \begin{array}{ll}
    path(i, d) = j, & \forall j~s.t.~ dis_e(i,j,d)=dis_n(i,d)\\
   \end{array}
 \end{equation}

We use a dynamic programming based method to solve distance
$dis_e(i, j, d)$, $dis_n(i, d)$ and $path(i, d)$, as shown in
Algorithm \ref{alg:dp}.

\begin{figure}[tb]
\centering
\includegraphics[width=0.44\textwidth]{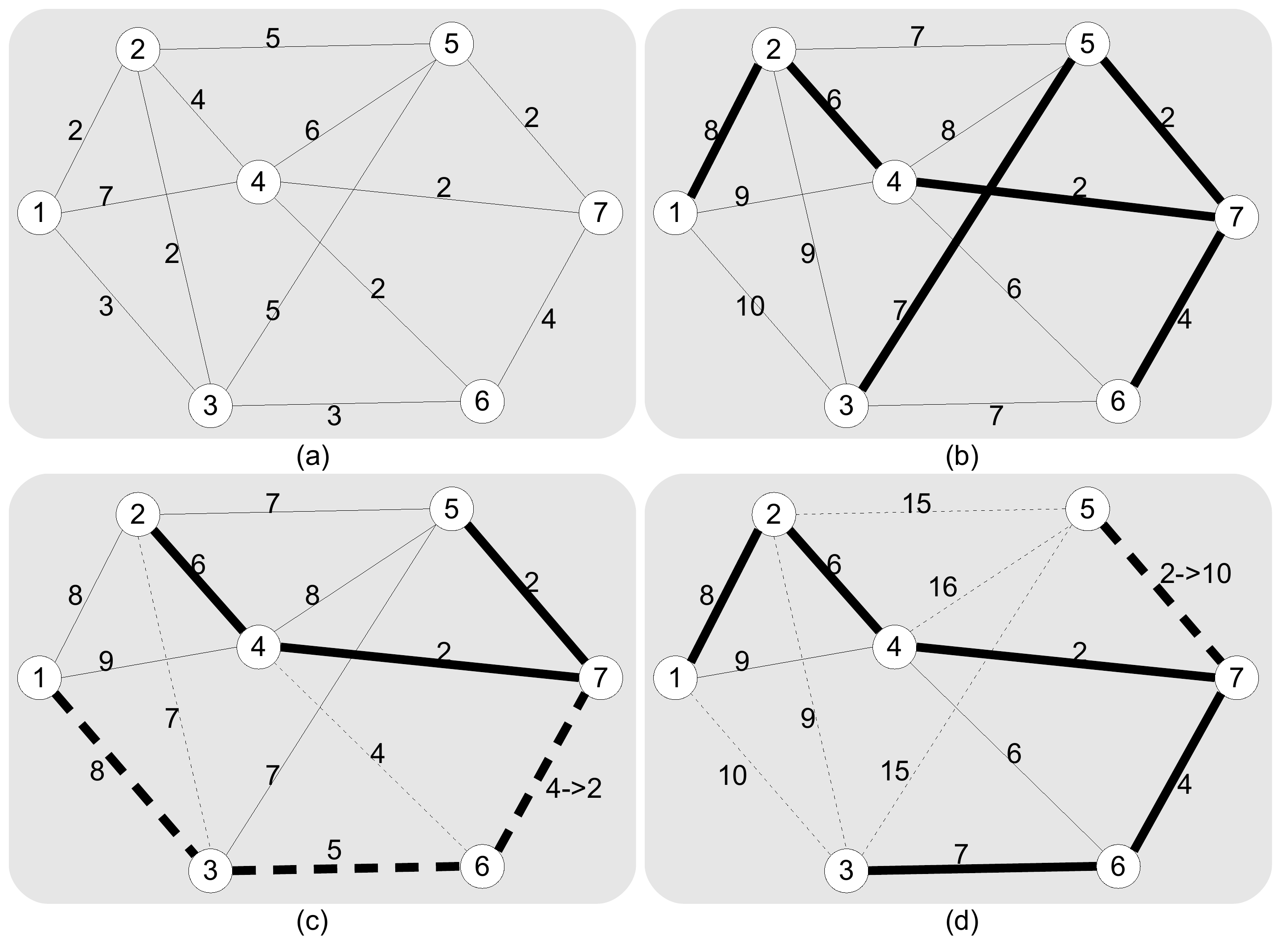}
\caption{~A simple example of paths allocation with seven
switches.~(a)Initial network, the value on each edge $e_{ij}$ is
$t_{ij}$. ~(b)After $InitSolve(7)$, the value on each edge $e_{ij}$
is $dis_e(i, j, 7)$ and each bold edge $e_{ij}$ means $path(i,
7)=j$.~(c)Compare with (b), $t_{67}$ decreases from 4 to 2, update
some edges(labeled as dotted arrows). ~(d)Compare with (b), $t_{57}$
increases from 2 to 10, update some edges(labeled as dotted
arrows).} \label{path}
\end{figure}

\begin{algorithm}[!tb]
\caption{$InitSolve(d)$} \label{alg:dp}
\begin{algorithmic}[1]
 \STATE //Given $d$, solve all $dis_e(i, j, d)$ and $dis_n(i, d)$;
 \STATE Initialize all $D(i, j, d) \leftarrow M$;
 \FORALL{$k \in Pre(d)$}
   \STATE $dis_e(k, d, d) \leftarrow t_{kd};$
   \STATE $dis_n(k, d) \leftarrow t_{kd};$
 \ENDFOR
 \FOR{$i=d-1$ to $1$}
     \FORALL{$j \in Post(i)$}
       \STATE $dis_e(i, j, d) \leftarrow t_{ij} + dis_n(j, d);$
     \ENDFOR
     \STATE $dis_n(i, d) \leftarrow min_j~ dis_e(i, j, d), \forall j\in Post(i)$;
     \STATE $path(i, d) \leftarrow j$;
 \ENDFOR
\end{algorithmic}
\end{algorithm}

\begin{Theorem}
The required time for Algorithm $InitSolve()$ is at most $O(|E|)$.
The run time to solve all the nodes is bounded by $O(|V|\cdot|E|)$.
\end{Theorem}

If $t(i, j)$ changes, instead resolving all the $dis_e(i, j, d)$ and
$dis_n(i, d)$, we can effectively update them. If $t(i, j)$
decreases, we use Algorithm \ref{alg:decrease}, otherwise we use
Algorithm \ref{alg:increase}.

\begin{algorithm}[!tb]
\caption{$DecreaseUpdate(i, j, \Delta t, d)$} \label{alg:decrease}
\begin{algorithmic}[1]
   \STATE //Update when $t_{ij}$ change to $(t_{ij}-\Delta t)$;
   \STATE $t_{ij} \leftarrow (t_{ij}- \Delta t)$;
   \STATE queue q.push($e_{ij}$);
   \WHILE{q is not empty}
     \STATE $e_{ab} \leftarrow$ q.pop();
     \STATE $dis_e(a,b,d) \leftarrow t_{ab}+dis_n(b,d)$;
     \IF{$t_{ab}+dis_n(b,d) < dis_n(a,d)$}
       \STATE $dis_n(a,d) \leftarrow t_{ab}+dis_n(b,d)$;
       \STATE $path(a,d) \leftarrow b$;
       \STATE q.push($e_{pa}$), $\forall p\in Pre(a)$;
     \ENDIF
   \ENDWHILE
\end{algorithmic}
\end{algorithm}

\begin{algorithm}[!tb]
\caption{$IncreaseUpdate(i, j, \Delta t, d)$} \label{alg:increase}
\begin{algorithmic}[1]
   \STATE //Update when $t_{ij}$ change to $(t_{ij}+\Delta t)$;
   \STATE $t_{ij} \leftarrow (t_{ij}+ \Delta t)$;
   \STATE queue q.push($e_{ij}$);
   \WHILE{q is not empty}
     \STATE $e_{ab} \leftarrow$ q.pop();
     \STATE $dis_e(a,b,d) \leftarrow t_{ab}+dis_n(b,d)$;
     \IF{$PATH[a][d]=b$}
       \STATE Find $k \in Post(a)$ to minimize $dis_n(k,d)+t_{ak}$;
       \STATE $dis_n(a,d)\leftarrow dis_n(k,d)+t_{ak}$;
       \STATE $path(a,d)\leftarrow k$;
       \STATE q.push($e_{pa}$), $\forall p\in Pre(a)$;
     \ENDIF
   \ENDWHILE
\end{algorithmic}
\end{algorithm}

We consider a simple paths allocations as shown in Fig.\ref{path}. A
SCG with seven switches is shown in (a), the value on each edge
$e_{ij}$ is initial $t_{ij}$. Using Algorithm \ref{alg:dp} setting
$d=7$, we can solve each $dis_e(i, j, 7)$(labeled on each edge in
(b)). If $t_{67}$ decreases from 4 to 2, we use Algorithm
\ref{alg:decrease} to update some $dis_e(i, j, 7)$ and $dis_n(i,
7)$. As shown in (c), queue $q$ pushes edges $e_{67}, e_{36},
e_{46}, e_{13}, e_{23}$ one by one(labeled as dotted arrows). And
$path_{3, 7}$ changes from 5 to 6 and $path_{1, 7}$ changes from 2
to 3. If $t_{57}$ increases from 2 to 10, we use Algorithm
\ref{alg:increase} to update $dis_e(i, j, 7)$ and $dis_n(i, 7)$. As
shown in (d), queue $q$ pushes edges $e_{37}, e_{35}, e_{45},
e_{13}, e_{23}$ one by one(labeled as dotted arrows). And $path_{3,
7}$ changes from 5 to 6.

\section{Experimental Results}
We implemented our algorithm in the C++ programming language and
executed on a Linux machine with a 3.0GHz CPU and 1GB Memory. During
floorplanning we use hMetis\cite{DAC97Karypis}, an efficient
hierarchical graph partitioning tool.

\subsection{Power Model}
NoC power consumption consists of two parts: power consumed by
interconnects and power consumed by switches
 %\footnote{Here we ignore power consumption caused by network interfaces.}.
For each network link $e$, we assume $P_e$ represents bit energy on
link $e$ and the corresponding switches. $P_e =  P_l + P_s$, where
$P_l$ and $P_s$ are bit energy on interconnects and switches,
respectively. Power consumption is $P=P_e \times f$, where $f$
represents communication requirements passing the link and the
corresponding switch. We use Orion\cite{02ORION} as power simulator.
Table II gives the switch bit energy in $0.18um$ technology and
Table III gives the power model of links.

\begin{table}[tb]
\centering \caption{Power Model of Switch }
 \footnotesize
\begin{tabular}{|c|c|c|c|c|c|c|c|}
 \hline
 ports & 2 & 3 & 4 & 5 & 6 & 7 & 8\\
 \hline
 (pJ/bit) & 0.22 & 0.33 & 0.44 & 0.55 & 0.66 & 0.78 & 0.90\\
 \hline
\end{tabular}
\end{table}

\begin{table}[tb]
\centering \caption{Power Model of Interconnects }
 \footnotesize
\begin{tabular}{|c|c|c|c|c|c|}
 \hline
 Wire length(mm) & 1 & 4 & 8 & 12 & 16\\
 \hline
 (pJ/bit) & 0.6 & 2.4 & 4.8 & 7.2 & 9.6\\
 \hline
\end{tabular}
\end{table}

\begin{table*}[tb]\label{table:result}
 \footnotesize
 \centering \caption{The Consumption Between the PDF and the PBF}
\begin{tabular}{|c|c|c|c|c|c|c|c|c|c|c|}
 \hline \hline
 Benchmark & V\# & E\# & Part\# & \multicolumn{2}{|c}{Power(mW)} & \multicolumn{2}{|c}{Hops}
           &\multicolumn{2}{|c|}{W.S(\%)} & Time(s)\\
 \cline{5-11} & & & & PBF & ours & PBF & ours &PBF& ours & ours \\
 \hline
 MPEG4       & 12 & 13 & 3 & 25.9 & 16.0 & 1.17 & 1.0 & 12.25 & 16.43 &  13.86 \\
             &    &    & 4 & 24.3 & 14.1 & 1.25 & 1.041 & 7.63 & 16.43 &  15.07 \\
 \hline
 MWD         & 12 & 12 & 3 & 3.05 & 3.08 & 1.33 & 1.33 & 12.22 & 11.82 & 13.37\\
             &    &    & 4 & 3.19 & 3.02 & 1.25 & 1.25 & 12.22 & 12.22 & 15.46\\
 \hline
 VOPD        & 12 & 14 & 3 & 7.43 & 6.12 & 1.0 & 1.0 & 12.16 & 13.54 & 14.54 \\
             &    &    & 4 & 7.62 & 6.59 & 1.0 & 1.15 & 12.17 & 13.85 & 17.32 \\
 \hline
 263decmp3dec & 14 & 15 & 3 & 4.96  & 3.92 & 1.0 & 1.0 & 14.24 & 13.44 & 23.78 \\
              &    &    & 4 & 7.86 & 4.35 & 1.25 & 1.0 & 13.59 & 14.50 & 24.96 \\
 \hline
 263encmp3dec & 12 & 12 & 3 & 24.7 & 19.2 & 1.0 & 1.0 & 6.06 & 8.82 & 13.19 \\
              &    &    & 4 & 58.6 & 19.2 & 1.0 & 1.0 & 9.58 & 9.58 & 15.42 \\
 \hline
 mp3encmp3dec & 13 & 13 & 3 & 8.4 & 4.4 & 1.0 & 1.0 & 15.23 & 17.60 & 20.29 \\
              &    &    & 4 & 11.2 & 8.6 & 1.0 & 1.0 & 15.23 & 15.24 & 21.0\\
 \hline
 D\_38\_tvopd & 38 & 47 & 3 & 12.7 & 8.2 & 1.33 & 1.33 & 15.1 & 24.5 & 92.7 \\
              &    &    & 4 & 12.3 & 6.8 & 1.44 & 1.4 & 14.7 & 22.60 & 104.0 \\
 \hline
 Avg & -&-&-& 15.16 & 8.83 & 1.14 & 1.11 & 12.31 & 13.92 &28.93\\
 \hline
 Diff & -&-&-&-&-41.8\% & - & -2.6\% & - & - & - \\
\hline \hline
\end{tabular}
\end{table*}

\subsection{Results and discussion}
\begin{figure}[tb]
\centering
\includegraphics[width=0.2\textwidth]{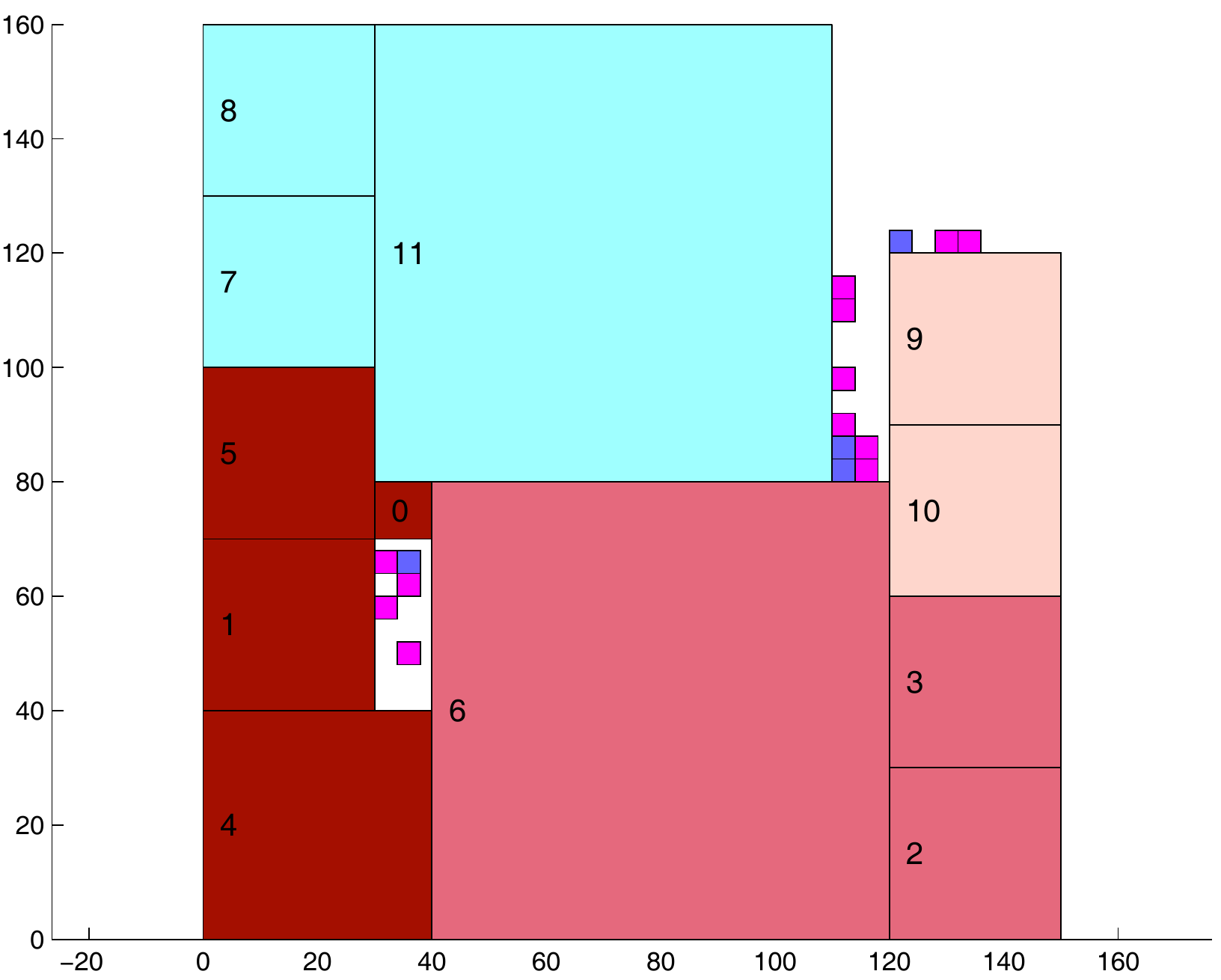}
\includegraphics[width=0.2\textwidth]{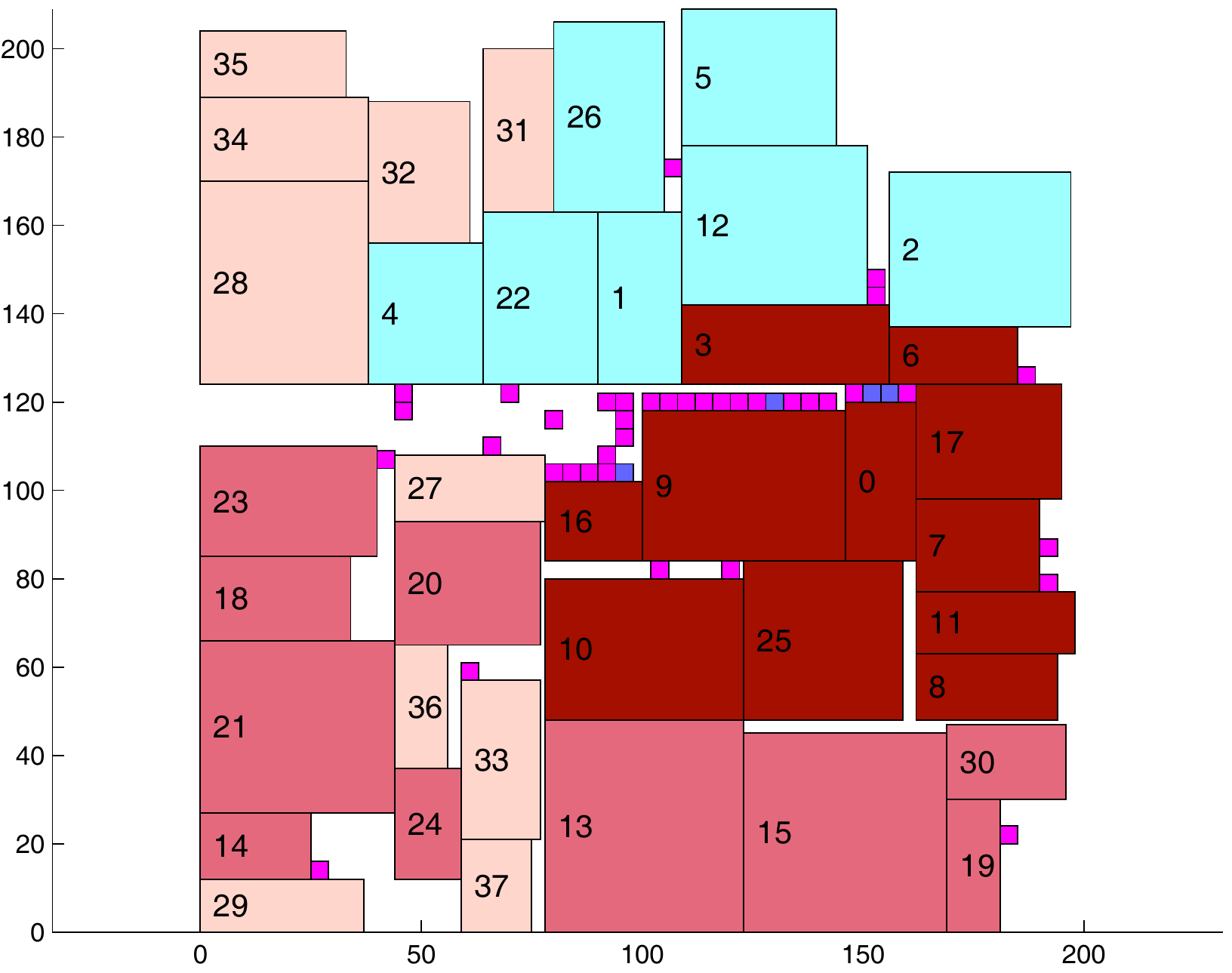}
\caption{Experimental results of 263decmp3dec and D\_38\_tvopd with
four clusters.}\label{result}
\end{figure}

We have applied our topology generation procedure to three sets of
benchmarks. The first set of benchmarks are several video processing
applications obtained from \cite{ieee05Bertozzi}: MPEG4, MWD and
VOPD. The next set of benchmarks are obtained from
\cite{ieee06Srinivasan}: 263decmp3dec, 263encmp3dec and
mp3encmp3dec. The last benchmark is obtained from
\cite{ASPDAC09MURALI}: D\_38\_tvopd. Fig.\ref{result} shows two
floorplan generated for the 263decmp3dec and D\_38\_tvopd benchmark.

We performed experiments to evaluate our topology generation
algorithm. For comparison, we have also generated another approach
PBF, which is similar to the min-cut based algorithm presented in
\cite{ICCAD06MURALI}. In PBF, partition is solved only before
floorplanning. Table IV shows comparisons between our experimental
results and PBF. The column Power means the actual power consumption
and column Hops means average number of hops. Our method can save
41.8\% of power and 2.6\% of hops number. For test cases that have
more communication requirements, such as 263encmp3dec, our algorithm
can save much more power(reduce power consumption from 58.6 mW to
19.2 mW). The column W.S means the white spaces and column Time is
run time. The white space of our method increases from 12.31\% to
13.92\% and run time is reasonable. Since power saving is the most
important concern, the deteriorating is acceptable.

\begin{table}[tb]\label{table:result2}
 \footnotesize
 \centering \caption{Comparison for Fault Tolerant}
\begin{tabular}{|c|c|c|c|c|c|c|}
 \hline \hline
  & V\# & Flow\# & Update\# & \multicolumn{2}{|c|}{Run Time(s)} & Diff\\
 \cline{5-6}
 & & & & DSP & ours & \\
 \hline
 t\_01  & 20 & 34 & 20 & 0.024 & 0.008 & -66.7\%\\
 \hline
 t\_02  & 100 & 130 & 30 & 0.604 & 0.016 & -97.4\%\\
 \hline
 t\_03  & 300 & 457 & 50 & 20.35 & 0.08 & -99.6\%\\
\hline \hline
\end{tabular}
\end{table}

We further demonstrated the effectiveness of Algorithm
\ref{alg:decrease} and Algorithm \ref{alg:increase}. To update
routing when link cost changes, we performed another contrastive
approach DSP. DSP re-solves all the distances of flows by Dijkstra's
shortest path algorithm\cite{book:flow}. We have applied another set
of test cases: t\_01, t\_02 and t\_03. For each case, table V
reports the number of nodes V\#, flow number and update times
Update\#. We can see that our updating algorithm can save lots of
run time: t\_01 saves 66.7\%, t\_02 saves 97.4\% and t\_03 can save
99.6\%.

%Table V shows comparisons between our method and DSP. The column V\#
%means the number of nodes in the case, column Flow\# means the
%number of flows and column Update\# means update times. Our method
%can save lots of run time and is effective.

\section{CONCLUSIONS}

We have proposed a two phases framework to solve topology synthesis
for NoCs: phase one is partition driven floorplanning; phase two is
switches insertion, network interfaces insertion and paths
allocations to minimize power consumption. Experimental results have
shown that our framework is effective and can save power consumption
by 41.8\%.

%*********************************************************
%                 
%*********************************************************

\end{document}